\documentclass{emulateapj}
\usepackage{graphicx}

\newcommand{\Cu}[7]{\mbox{${#1}{#2}^{#3}\,^#4{#5}^{\rm #6}_{#7}$}}
\newcommand{\Teff}{T_{\rm eff}}
\newcommand{\kms}{km s$^{-1}$}
\newcommand{\SH}{$S_{\!\!\rm H}$}
\newcommand{\eps}[1]{\log{\varepsilon}_\sun {\rm #1}}
\newcommand{\afe}{[$\alpha$/Fe]}

\slugcomment{Draft version 20150106}
\shorttitle{Non-LTE analysis of neutral copper}
\shortauthors{Yan et al.}

\begin{document}

\title{Non-LTE analysis of neutral copper in the late-type metal-poor stars \altaffilmark{$\star$}}

\author{H.L. Yan \altaffilmark{1,2}, J.R. Shi \altaffilmark{1}, and G. Zhao \altaffilmark{1}}

\altaffiltext{$\star$}{Based on the observations collected at the Germany-Spainish Astronomical Center, Calar Alto, Spain.}
\altaffiltext{1}{Key Laboratory of Optical Astronomy, National Astronomical Observatories, Chinese Academy of Sciences,
                 Beijing 100012, China;
                 \email{sjr@nao.cas.cn}}
\altaffiltext{2}{University of Chinese Academy of Sciences,
                 Beijing 100049, China}

\date{Received date / Accepted date}

\begin{abstract}
We investigated the copper abundances for $64$ late-type stars in the Galactic disk and halo with effective temperatures from $5400$ K to $6700$ K and [Fe/H] from $-1.88$ to $-0.17$. For the first time, the copper abundances are derived using both local thermodynamic equilibrium (LTE) and non-local thermodynamic equilibrium (non-LTE) calculations. High resolution ($R > 40,000$), high signal-to-noise ratio ($S/N > 100$) spectra from the FOCES spectrograph are used. The atmospheric models are calculated based on the MAFAGS opacity sampling code. All the abundances are derived using the spectrum synthesis methods. Our results indicate that the non-LTE effects of copper are important for metal-poor stars, showing a departure of $\sim 0.17$ dex at the metallicity $\sim -1.5$. We also find that the copper abundances derived from non-LTE calculations are enhanced compared with those from LTE. The enhancements show clear dependence on the metallicity, which gradually increase with decreasing [Fe/H] for our program stars, leading to a flatter distribution of [Cu/Fe] with [Fe/H] than previous work. There is a hint that the thick- and thin-disk stars have different behaviors in [Cu/Fe], and a bending for disk stars may exist.
\end{abstract}


\section{INTRODUCTION}

The investigation of the elemental abundance traces the evolution of the Galaxy. The chemical history of the Galaxy is dominated by the processes of nucleosynthesis in each generation of stars. Based on the observational behaviors of elemental abundances in stars with different metallicities, one can not only look back the Galactic chemical enrichment history, but also constrain the theoretical evolutionary models of our Galaxy. Copper is of particular interest among the iron-peak elements, because firstly it has unique evolutionary trend as the [Fe/H] varies from extremely metal-poor to the solar abundance, and secondly this element is thought to be synthesized by several possible nucleosynthesis scenarios, yet the contributions of these scenarios are still in dispute.

From the observational point of view, although it was \citet{Coh78,Coh79,Coh80} who gave the early glance at the Galactic copper abundances (they investigated $27$ red giants in $7$ globular clusters), and who first suggested a decreasing trend of [Cu/Fe] with decreasing [Fe/H], the evolutionary trend of [Cu/Fe] was not established firmly until \citet{Sne91}. In a series of work \citep{Gra88,Sne88,Sne91}, the authors derived the copper abundances for a large sample of stars in the Galactic disk and halo. Their results clearly showed a linear trend of [Cu/Fe], which increases towards the higher metallicity. This trend was partly confirmed by \citet{Mis02} and \citet{Sim03}, the former authors investigated an expanded sample of disk and halo stars with wide range metallicity, while the latter ones managed to measure the copper abundances for $117$ giants in $10$ globular clusters. Furthermore, both work indicated a flat plateau of [Cu/Fe] at the metal-poor end, which is roughly [Cu/Fe] $\approx -0.75$ at [Fe/H] $< -1.5$. Some contemporaneous studies on single ultra-metal-poor stars seemed consistent with this flat distribution \citep[e.g.,][]{Wes00,Cow02,Sne03}. However, using the near-UV lines, \citet{Bih04} and \citet{Lai08} suggested the plateau should be around $-1$ instead of $-0.75$. \citet{Red03,Red06} investigated a large sample of disk stars, and they found little variation of [Cu/Fe] with [Fe/H] in the metallicity range [Fe/H] $> -0.8$, where the values of [Cu/Fe] show no evident divergence with respect to the solar copper abundance. Thus, a slight S-shape of [Cu/Fe] as a function of [Fe/H] can be seen by overlapping the data from different authors \citep[Figure 1]{Bih04}. Although the stars investigated by a great quantity of work followed the Galactic general S-shape \citep[e.g.,][]{Pro00,She01,She03,Coh08,Mis11}, several peculiar structures were still detected, such as Ursa Major moving Group (UMaG) \citep{Cas99}, Omega Centauri ($\omega$ Cen) \citep{Smi00,Cun02}, Sagittarius dwarf spheroidal galaxy (Sgr dSph) \citep{Mcw05,Mcw13}, and the halo sub-population \citep{Nis11}. Their discordant trends of [Cu/Fe] may imply different chemical evolutionary histories with respect to our Milky Way.

Theoretically, copper is thought to be produced in multiple astrophysical sites. The first one is the weak $s-$process which takes place in massive stars during the helium- and carbon-shell burning stage \citep{Sne91}. It is a secondary process that needs iron seeds from previous generations of stars, resulting in the linear dependence of [Cu/Fe] on [Fe/H]. \citet{Bis04} proposed a revised version of this scenario, suggesting $sr-$process dominated copper synthesis in massive stars instead of classical $s-$process. Additionally, simulations of Galactic chemical evolution (GCE) by \citet{Rom07} and \citet{Rom10} also support that massive stars contribute most of copper. However, \citet{Mat93} showed the second possibility by fitting GCE models to interpret observational data. They suggested that the main source of copper was type Ia supernovae (SNe Ia) instead of $s-$process (but they needed to increase the yields from SNe Ia by about an order of magnitude). This conclusion was subsequently supported by the work from \cite{Mis02} and \citet{Sim03}. The third mechanism comes from the constraint of the [Cu/Fe] plateau at low metallicity, which requires the contributions from primary explosive nucleosynthesis in type II supernovae (SNe II) \citep[e.g.,][]{Tim95}. However, the productions calculated from SNe II are model-dependent \citep{Woo95,Kob06,Nom06,Rom10}, resulting in discordant predicted fractions. Besides, large scatters in observational results also play a relevant role \citep[see the examples given by][]{Pig10}. The last source was the main $s-$process operating in the low- and intermediate-mass AGB stars, which was thought to contribute only about $5\%$ of the solar copper \citep{Tra04}.

So far, a number of analyses on copper abundance have been presented, covering the whole range of metallicity, from solar to the most extreme metal-poor stars, but none of them is carried out with the non-local thermodynamic equilibrium (non-LTE) calculations. This is partly because non-LTE calculations need reliable atomic model and detailed statistical equilibrium calculations, whereas neither of them is a simple job for copper. Despite the difficulties, there are at least two principal reasons that we have to put non-LTE analysis into perspective. First, recent studies on non-LTE effects have demonstrated that the non-LTE correction is large for some elements in metal-poor stars \citep[e.g.,][]{Bau97,Bau98,Zha98,Zha00,Geh04,Ber08,Shi04,Shi09}. Moreover, only a few lines can be used to perform copper abundance analysis when dealing with the metal-poor stars with relatively higher temperature, as there is little neutral copper in the atmospheres of such stars. The analysis has to rely on the two resonance lines at $3247$\,\AA\ and $3273$\,\AA, both of which may suffer large non-LTE effects \citep{Roe12,Roe14}. \citet{Bih04} and \citet{Bon10} also reported that the abundances derived from these two lines are inconsistent with those derived from other optical \ion{Cu}{1} lines.

In this paper, we aim at exploring the copper abundances for the sample stars in the metallicity range $-1.88 <$ [Fe/H] $< -0.17$ with a complete spectrum synthesis method based on level population calculated from the statistical equilibrium equations. In Section 2, we will briefly introduce some key information about the observations. The fundamental work for abundance analysis, i.e., atmospheric model, stellar parameters, and atomic line data are presented in Section 3. The 4th section will provide details of non-LTE calculations including the atomic model. In Section 5 we will show the final results and error analysis. The discussions and conclusions are presented in Section 6 and Section 7, respectively.

\section{OBSERVATIONS}

The sample stars investigated in this paper have already been discussed by \citet{Geh04,Geh06}. Here, we simply list the key features of the sample and the observations. More details can be found in aforementioned papers.

\begin{itemize}
\item[--] The observations were carried out on the $2.2$ m telescope located at Calar Alto Observatory from the year of $1999$ to $2003$. The FOCES \'{e}chelle spectrograph was used to obtain the high-resolution spectra, providing $97$ spectral orders in total that started at $3700$\,\AA\ and ended at $9800$\,\AA.

\item[--] The detector was a CCD chip with $2048\times2048$ pixels, and the size for each pixel is $24$ $\mu$m. A two-pixel bin results in a $\sim 40,000$ resolution power ($R$).

\item[--] The total exposure time is divided into more than two exposures. The final combined spectra show a signal-to-noise ratio ($S/N$) that is higher than $100$.

\item[--] The spectra were reduced from the program designed for FOCES spectrograph \citep{Pfe98}, which worked under the IDL environment. Cosmic rays and bad pixels were removed by careful comparisons of the exposures from the same object. The instrumental response and background scatter light were also considered during the data reducing.
\end{itemize}

\section{FUNDAMENTAL WORK FOR ABUNDANCE ANALYSIS}

In this section, we briefly describe the methods and key information our studies based on, including the atmospheric model, the determinations of stellar parameters, the atomic line data, and the kinematic properties of our program stars.

\subsection{Model Atmosphere} \label{sec3-1}

Stellar atmospheric model is the foundation of the spectrum synthesis. Our work adopted the MAFAGS opacity sampling (OS) model. This code was developed by \citet{Gru04} and updated by \citet{Gru09} with the new iron atomic data computed by \citet{Kur09}. MAFAGS OS code describes a one-dimensional plane-parallel model with $80$ layers overall in the hydrostatic equilibrium state. The chemical homogeneity and local thermal equilibrium is assumed throughout the atmosphere. This model atmosphere was also applied in our previous work \citep{Mas11,Shi14}.

The comparison between MAFAGS and MARCS OS model has been performed by \citet{Shi14}. Although differences of the temperatures between those two models exist outside $\log \tau \simeq -4$ and inside $\log \tau \simeq 0.3$, both of these regions barely influence the synthetic copper spectral line profile.

\subsection{Stellar Parameters} \label{sec3-2}

For all of our program stars, we directly adopted the stellar parameters derived by \citet{Geh04,Geh06}. By fitting the theoretical profiles to the observational data, two Balmer lines were used to derive the effective temperatures ($\Teff$). The hydrogen broadening theory involved in the theoretical profile calculation was from \citet{Ali66}. The surface gravities ($\log g$) were obtained by $[g] = [Mass/Luminosity] + 4[\Teff]$, where the stellar mass and luminosity were evaluated with the help of evolutionary tracks \citep{Van00} and HIPPARCOS parallaxes. In addition, the microturbulence velocities ($\xi$) were determined simultaneously with the metallicities ([Fe/H]): the [Fe/H] determined from \ion{Fe}{2} line is supposed to be independent of the equivalent width. The final uncertainties in $\Teff$, $\log g$, [Fe/H], and $\xi$ were estimated to be $\pm80$ K, $\pm0.05$ dex, $\pm 0.05$ dex, and $\pm 0.1$ \kms, respectively.

\subsection{Atomic Line Data} \label{sec3-3}

We used a set of calibrated $\log gf$ values, each of which could reproduce the solar copper abundance independently, as presented by \citet{Shi14}. Furthermore, the van der Waals damping constants ($\log C_6$) for \ion{Cu}{1} were calculated according to \citet{Ans91,Ans95}. Five \ion{Cu}{1} lines can be seen in our FOCES spectra. For the evaluation of copper abundance, $5220.070$\,\AA\ and $5700.240$\,\AA\ are not good indicators, for both of them are badly blended in most of our program stars. As a result, the rest three lines are applied to our research, which are $5105.541$\,\AA, $5218.202$\,\AA, and $5782.132$\,\AA. The atomic line data of the three lines are listed in Table \ref{tab1}.

\begin{table}
\scriptsize
\begin{center}
\caption[1]{Atomic data of copper lines used in this work \label{tab1}}
\begin{tabular}{lr@{ $ - $ }lcccc}
\hline\hline\noalign{\smallskip}
$\lambda_{\rm air}$ & \multicolumn{2}{c}{Transition} & $E_{\rm low}$ & $\log gf$ & $\log C_6$ \\
 (\AA)              & \multicolumn{2}{c}{}           & (eV)          &           &            \\
\noalign{\smallskip}
\hline\noalign{\smallskip}
 $5105.541$ &  \Cu{4}{s}{2}{2}{D}{ }{5/2}  & \Cu{4}{p}{ }{2}{P}{o}{3/2} & $1.389$ & $-1.64$ & $-31.67$ \\
 $5218.202$ &  \Cu{4}{p}{ }{2}{P}{o}{3/2}  & \Cu{4}{d}{ }{2}{D}{ }{5/2} & $3.817$ & $+0.28$ & $-30.57$ \\
 $5782.132$ &  \Cu{4}{s}{2}{2}{D}{ }{3/2}  & \Cu{4}{p}{ }{2}{P}{o}{1/2} & $1.642$ & $-1.89$ & $-31.66$ \\
\noalign{\smallskip} \hline
\end{tabular}
\end{center}
\textbf{Notes.} The $\log gf$ values were rectified from the non-LTE solar spectrum fitting, and the van der Waals damping constants ($\log C_6$) were calculated according to \citet{Ans91,Ans95} \\
\end{table}

\subsection{Population and Kinematic Properties} \label{sec3-4}

Similar to the stellar parameters, we also adopted the population identified by \citet{Geh04,Geh06} for our program stars, which was based on the kinematic features, stellar ages, [Al/Mg] and [Mg/Fe] ratios. Most of the stars were classified as thin-disk, thick-disk, and halo population, whereas the rest were the stars with peculiarities.

\section{Non-LTE CALCULATIONS}

LTE assumption provides us a simple way to calculate the population of each energy level and number densities of different ionization stages for a given element, while non-LTE calculations require to solve the detailed statistical equilibrium equations. Thus, a reliable atomic model of copper is indispensable.

The atomic model of copper has been described in the previous paper \citep{Shi14} and the Grotrian diagram of the model can also be seen there. We modeled the copper atom with $17$ orbits, $97$ energy levels ($96$ states for \ion{Cu}{1} and the ground state for \ion{Cu}{2}) and $1089$ transitions, and the fine structure for the levels with low excitation energy was also included. The atomic data of such complex structure are obtained from both laboratory measurements (NIST\footnote{http://www.physics.nist.gov/} database, \citeauthor{Sug90} \citeyear{Sug90}) and theoretical calculations \citep{Liu14}. In addition, the excitation and ionization caused by inelastic collisions were also considered. The data for collisions with neutral hydrogen were obtained based on the \citeauthor{Dra68} formula \citep{Dra68,Dra69} presented by \cite{Ste84}, and we decreased the collisional rates by an order of magnitude (\SH $= 0.1$) under the suggestion of \citet{Shi14}. The excitation and ionization caused by inelastic collisions with electrons are calculated according to a number of theoretical work \citep{Van62,All73,Sea62}. We used a revised DETAIL program \citep{But85} with accelerated lambda iteration method to perform the statistical equilibrium calculations.

\begin{figure}
\epsscale{1.2}
\plotone{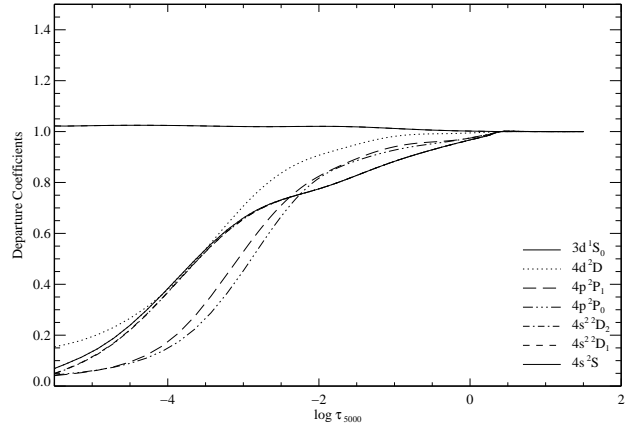}
\caption{The departure coefficients ($b_i$) for selected energy levels (listed in the figure) as a function of continuum optical depth at $5000$\,\AA\ for the model atmosphere of HD\,$59984$. The collision with neutral hydrogen was scaled by a factor of 0.1 in light of \citet{Shi14}. \\ \label{fig1}}
\end{figure}

In Figure \ref{fig1}, we present how the departure coefficients ($b_i=n_i^{\rm non-LTE}/n_i^{\rm LTE}$) of the selected levels vary with the continuum optical depth at $5000$\,\AA\ ($\log \tau_{\rm 5000}$) for the model atmosphere of HD\,$59984$ , where $b_i$ is the ratio of the number density of non-LTE ($n_i^{\rm non-LTE}$) to that of LTE ($n_i^{\rm LTE}$). HD\,$59984$ is a typical star with moderate temperature and metallicity among our sample, which is randomly selected as an example for the convenience of discussion. The departure coefficients for \ion{Cu}{1} important levels and \ion{Cu}{2} ground state are shown in the figure. It also shows that the number densities of these levels begin to underpopulate outside the layers with $\log \tau_{\rm 5000} \sim 0.5$ due to the overionization.

\section{RESULTS}

\subsection{Spectral Line Synthesis} \label{sec5-1}

\begin{figure}
\epsscale{1.2}
\plotone{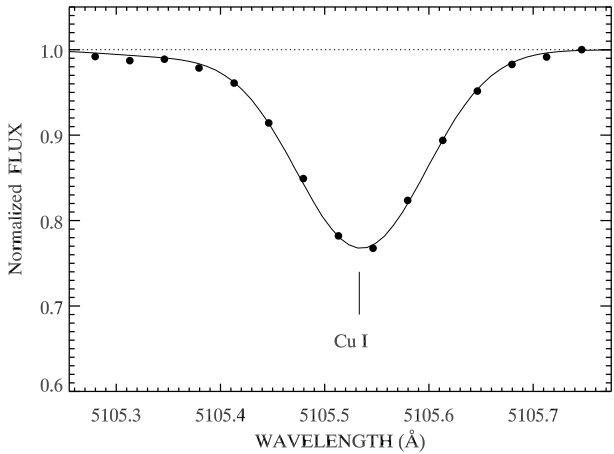}
\caption{The synthetic profile of the \ion{Cu}{1} $5105$\,\AA\ line for HD\,$59984$. The observed spectrum and theoretical synthesis are represented by filled circles and solid line, respectively.\\ \label{fig2}}
\end{figure}

In our analysis, the transitions of the hyperfine structure (HFS) were calculated according to \citet{Bie76} with RS coupling method. For the lines we used, if the intervals of HFS components are within $1$ m\AA, we combined them together. The adopted solar copper abundance is the value derived from the meteorites, which is $\eps {(Cu)} = 4.25$ \citep{Lod09}, and the oscillator strengths were also calibrated based on this value as described in aforementioned section. Additionally, the ratio between two copper isotopes ($^{63}$Cu and $^{65}$Cu) was assumed to be $0.69:0.31$ \citep{Asp09}. An IDL based program SIU was used to perform the line formation in our abundance determinations, which was developed by \citet{Ree91}. In our analysis, the broadenings caused by the macroturbulence, rotation, and instrument were treated as one single Gauss broadening factor, being convolved with the synthetic spectra to fit the observed line profile. The comparison between the synthetic and observed line profile at $5105$\,\AA\ for HD\,$59984$ is shown in Figure \ref{fig2} as an example. The observed spectrum and theoretical synthesis are represented by filled circles and solid line, respectively. The uncertainty of our line profile synthesis is less than $0.02$ dex.

\subsection{Copper Abundances and Error Analysis} \label{sec5-2}

The copper abundances are derived successfully for $60$ stars in our sample, and the selected \ion{Cu}{1} lines are too weak to rely on for the rest four stars, which are HD\,$241253$, HD\,$233511$, G\,$119-32$, and BD\,$+20^{\circ}\,2594$. The derived abundances with both LTE and non-LTE calculations are listed in Table \ref{tab2}, where the results of each individual line are also presented. The final abundance of each star is given by calculating the arithmetical mean value of every line used in the analysis.

\begin{figure*}
\epsscale{1.2}
\plotone{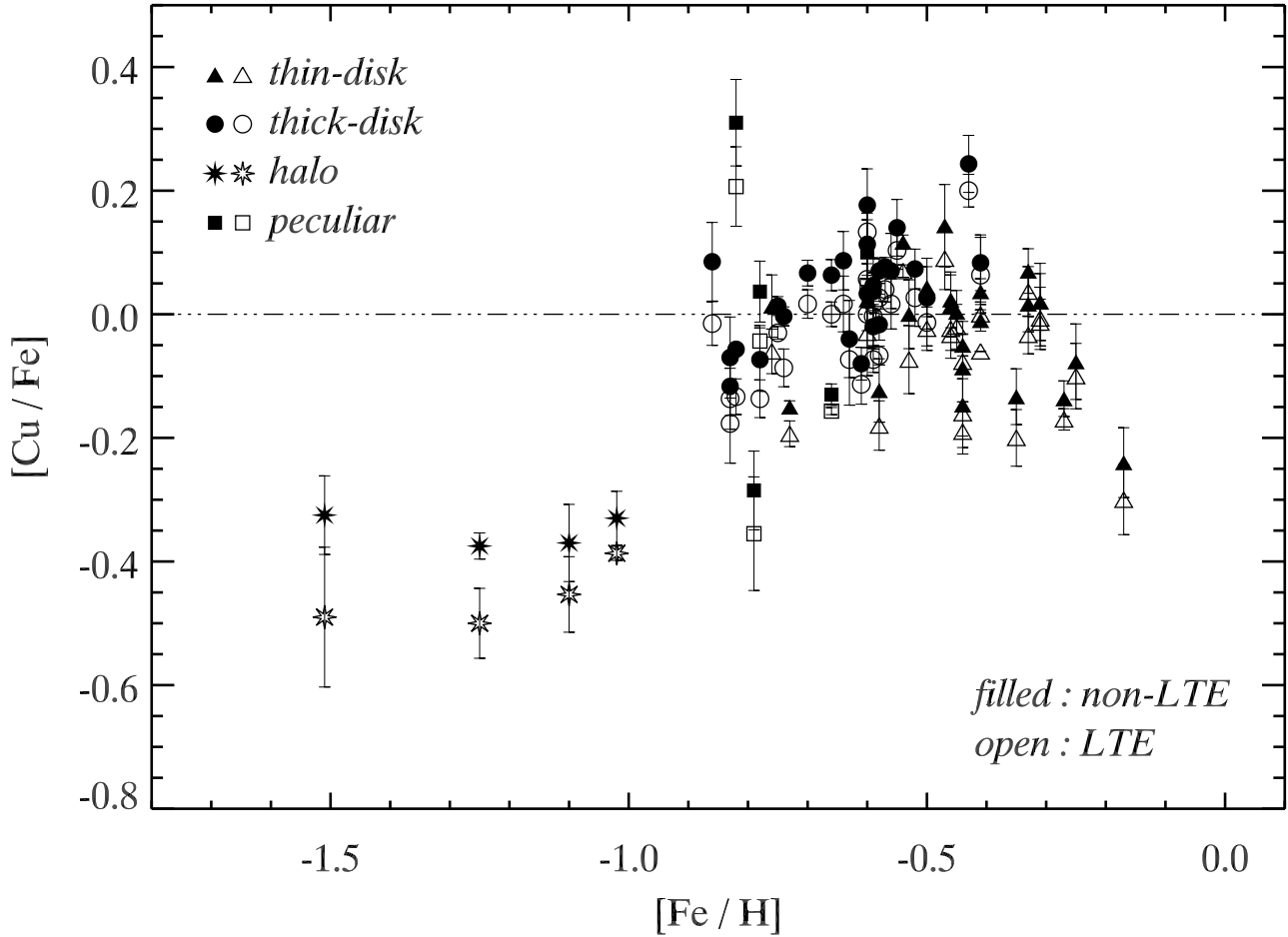}
\caption{Abundance ratios [Cu/Fe] as a function of [Fe/H] for our program stars, where non-LTE and LTE results are represented by filled and open symbols, respectively. Furthermore, symbols with different shapes represent stars from different populations, which are: triangle -- thin disk, circle -- thick disk, star -- halo, square -- objects with peculiarities. The errors are evaluated by computing the standard deviations of the abundances derived from different spectral lines.\\ \label{fig3}}
\end{figure*}

\begin{figure}
\epsscale{1.2}
\plotone{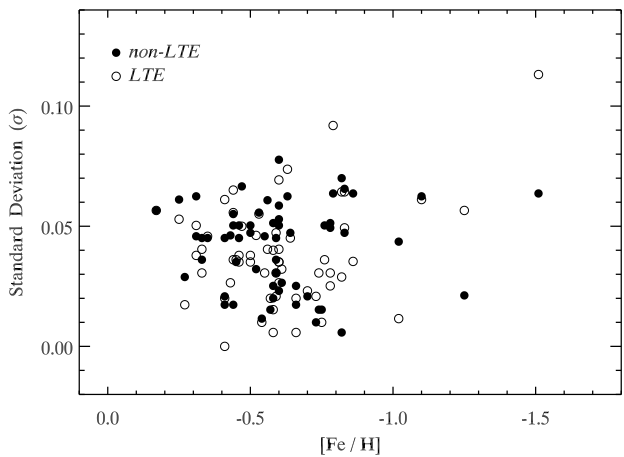}
\caption{The distribution of standard deviation as a function of metallicity, where filled and open circles represent non-LTE and LTE calculations, respectively. The stars whose copper abundances were derived by only one single line were not plotted in the figure.\\ \label{fig4}}
\end{figure}

The trend of [Cu/Fe] to [Fe/H] for our program stars is shown in Figure \ref{fig3}. The errors in the figure are evaluated by computing the standard deviations of the abundances derived from different spectral lines. Our results do not show large abundance discrepancy between different lines. We present the errors as a function of [Fe/H] in Figure \ref{fig4}. In LTE calculations, the standard deviations vary from $0.01$ to $0.11$, slightly larger than those in non-LTE, which are between $0.01$ and $0.08$. Furthermore, the LTE standard deviations become large at the metal-poor end, while the non-LTE ones remains stable. Both of them have a mean value that are around $0.04$. The errors caused by the uncertainties of the stellar parameters are estimated for HD\,$59984$, and the resulting effects in [Cu/Fe] are $\pm 0.07$, $<0.01$, $\pm 0.05$, and $<0.01$ dex for the typical uncertainties in $\Teff$, $\log g$, [Fe/H], and $\xi$, respectively.

\begin{figure}
\resizebox{8.8cm}{!}{\includegraphics{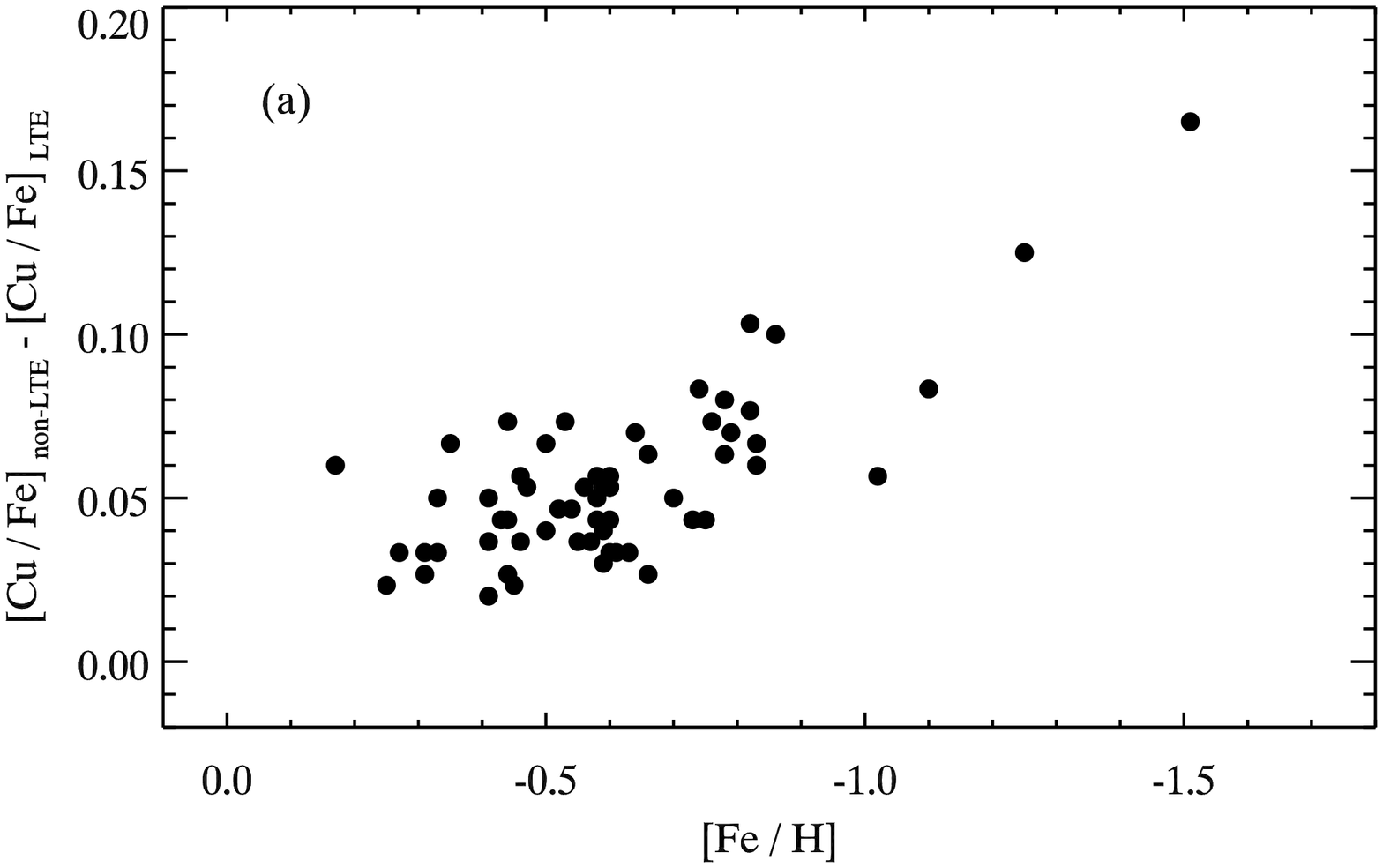}}\\[0.0cm]
\resizebox{8.8cm}{!}{\includegraphics{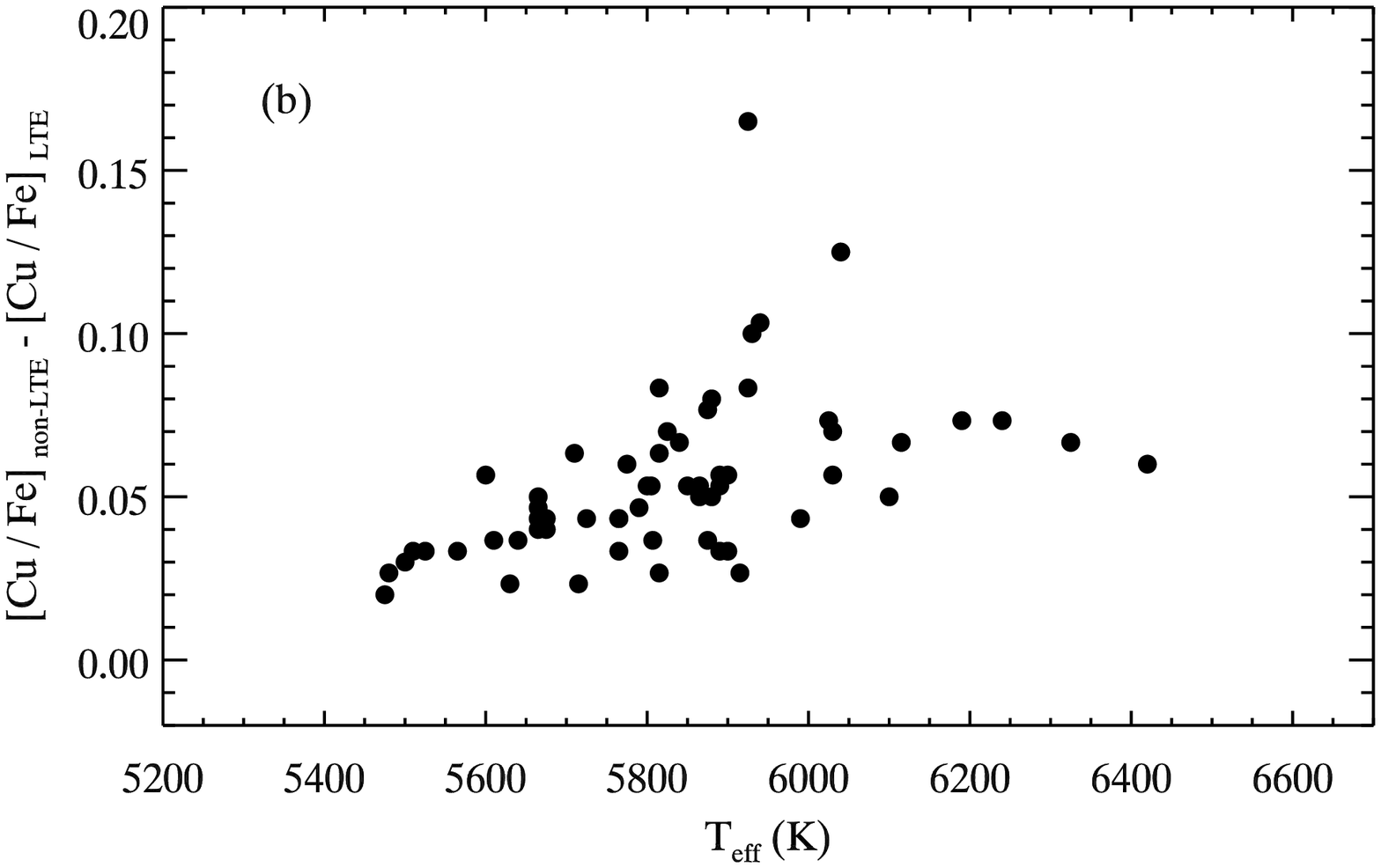}}\\[0.0cm]
\resizebox{8.8cm}{!}{\includegraphics{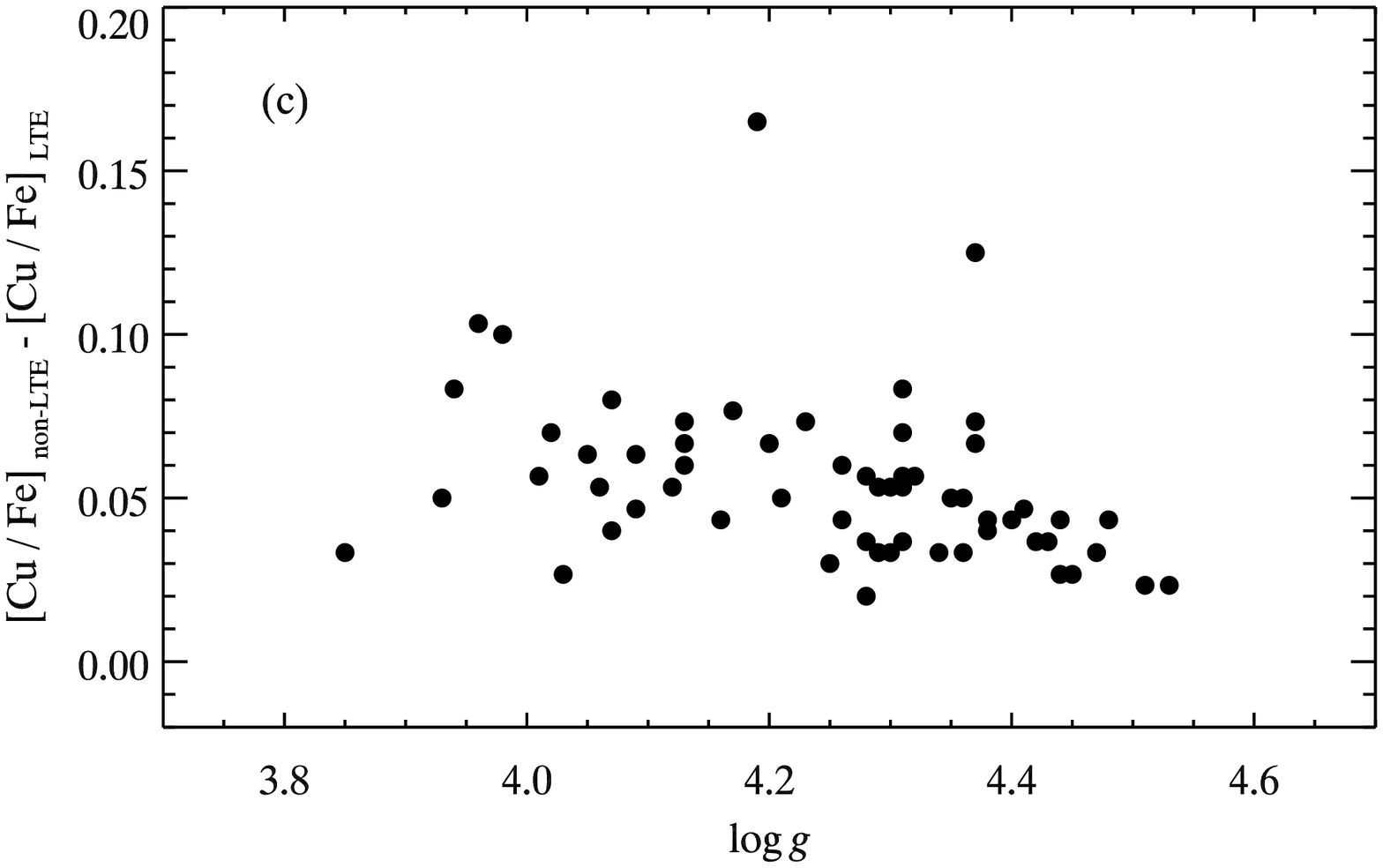}}\\[-0.3cm]
\caption[]{The differences in [Cu/Fe] between non-LTE and LTE for our program stars as a function of metallicity (a), effective temperature (b), and surface gravity (c).\\ \label{fig5}}
\end{figure}

The differences in [Cu/Fe] between non-LTE and LTE for our program stars as a function of metallicity, effective temperature, and surface gravity are plotted in Figure \ref{fig5}. The non-LTE departure shows clear dependence on the metallicity (see Figure \ref{fig5}a), which is gradually increases as [Fe/H] decreases in our program stars.

\subsection{Non-LTE Effects} \label{sec5-3}

The final results in Table \ref{tab2} show that the abundances derived from non-LTE calculations are larger than those from LTE ones for our program stars. The non-LTE correction for individual spectral line can reach $\sim +0.20$ dex for the stars with [Fe/H] $\sim -1.5$. The increasing of the copper abundance in non-LTE calculation is the consequence of the underpopulation of the lower energy levels (\Cu{4}{s}{2}{2}{D}{ }{ } and \Cu{4}{p}{ }{2}{P}{o}{ }), as shown in Figure \ref{fig1}. This deviation will lead to a underestimation of copper abundance by LTE analysis.

For each \ion{Cu}{1} line, the non-LTE effect is different, reflecting the properties of corresponding energy levels and associated transitions. The lines of $5105$\,\AA\ and $5782$\,\AA\ exhibit larger non-LTE effects compared with the weaker $5218$\,\AA. Taking HD\,$59984$ as an example again, the non-LTE corrections for the lines of $5105$\,\AA\ and $5782$\,\AA\ are $0.10$ and $0.09$ dex, respectively, while it is $0.06$ dex for the weaker $5218$\,\AA\ line. On one hand, despite the $5218$\,\AA\ line suffers less non-LTE effects than the other ones, it is usually blended by the line of $5217$\,\AA\ (\ion{Fe}{2} line), thus it is not a satisfactory indicator for the stars with low or moderate metallicity. On the other hand, since most abundance analyses are based on the two copper strong lines, one need to be aware of the non-LTE departures, especially for the metal-poor stars.

\subsection{Comparison with Other Work} \label{sec5-4}

\begin{figure}
\epsscale{1.2}
\plotone{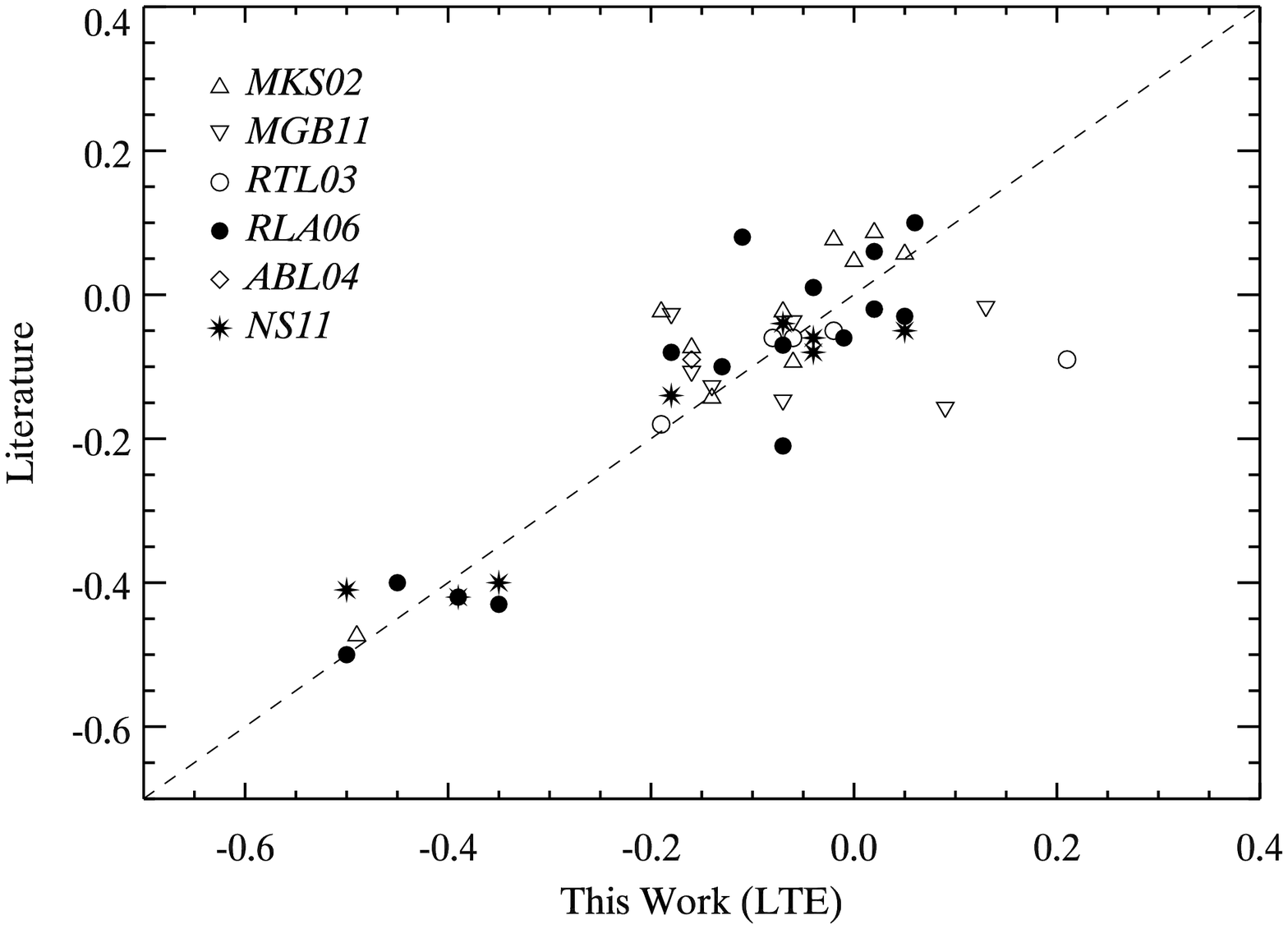}
\caption{Comparison of derived [Cu/Fe] in LTE analysis for the stars in common with the literature, including MKS02 \citep{Mis02}, MGB11 \citep{Mis11}, RTL03 \citep{Red03}, RLA06 \citep{Red06}, ABL04 \citep{All04}, and NS11 \citep{Nis11}. Corresponding symbols are annotated in the figure.\\ \label{fig6}}
\end{figure}

Numbers of studies for copper abundance have been carried out so far, allowing us to compare our results with those derived by several different groups based on LTE calculations. Comparing with the work published in recent years, we found $47$ stars are in common ($11$ of them are studied at least twice in different papers). In Figure \ref{fig6}, we compare our results derived from LTE calculations with those in the other work. In general, the LTE [Cu/Fe] values determined in our work are consistent with the different studies presented here, and no clear systematic deviation is found. We briefly discussed the details of the comparisons and the possible reasons of the large scatters.

\emph{\citet{Mis02,Mis11}}: \citeauthor{Mis02} studied copper abundances for $90$ metal-poor stars in the year of \citeyear{Mis02}, and later they investigated $172$ F to K dwarf stars and derived their copper abundances in \citeyear{Mis11}. All their observations were performed with $R = 42,000$ and $S/N > 100$. The copper abundances were derived from the $5105$\,\AA, $5218$\,\AA, and $5782$\,\AA\ lines, which were the same as those in our studies. The oscillator strengths adopted in their studies were from \citet{Gur89}. We have $13$ stars in common, and four of them are investigated in both of their papers. Since the derived [Cu/Fe] of these four stars varied a little bit for their two papers, we therefore adopted the latest values (derived in \citeyear{Mis11}) to perform the comparison. The resulting average difference of the $13$ stars is $\Delta[\rm Cu/Fe] = -0.01\pm0.12$. The main contribution of the relatively large scatter is from four stars, namely HD\,$22879$, HD\,$101177$, HD\,$108076$, and HD\,$218209$, as the [Cu/Fe] values derived from our work and theirs differ by more than $0.1$ dex for those stars. This is mainly due to the differences of the stellar parameters adopted in the two studies.

\emph{\citet{Red03,Red06}}: The authors performed an abundance analysis on a large sample of F and G dwarfs, containing thin-/thick-disk and halo stars. The copper abundances were determined from the $5105$\,\AA, $5218$\,\AA, and $5220$\,\AA\ lines. Their $\log gf$ values are similar to ours. From their samples, $21$ stars are found in common with ours, and the mean difference is $\Delta[\rm Cu/Fe] = 0.01\pm0.10$. Most of the stars show no large deviation except HD\,$200580$ and HD\,$107582$. The [Fe/H] values for these two stars adopted in their studies are different from those in ours. Taking [Fe/H] variances into account, the deviation can be perfectly removed.

\emph{\citet{All04}}: These authors analyzed a complete and comprehensive sample of $118$ stars with absolute magnitude brighter than $6.5$ and distance less than $14.5$ pc from the Sun. The spectral were obtained with $R = 50,000$ and $S/N > 150$. Compared with their sample, we have only one star in common. The difference of the [Cu/Fe] value is $-0.07$.

\emph{\citet{Nis11}}: \citeauthor{Nis11} carried out a series of elemental abundance analyses on a sample of $94$ dwarfs, most of which were identified as halo stars. They found the halo stars in the solar neighborhood fall into two distinct populations that can be separated by \afe. They used $5105$\,\AA, $5218$\,\AA, and $5782$\,\AA\ lines to derive copper abundances. The $\log gf$ values adopted in their analyses are similar to those used in ours. Our results are quite consistent with theirs. The average difference is $0.00\pm0.05$ for the $11$ stars in common.

\section{DISCUSSION}

\subsection{The Evolutionary Trend of [Cu/Fe] and Nucleosynthesis in the Galaxy} \label{sec6-1}

The observational trend of [X/Fe] as a function of [Fe/H] is a powerful tool to reveal the origins of elements and constrain the Galactic chemical evolutional model. The trend of copper for our Galaxy has been investigated in many papers, but all of the calculations are under LTE assumptions.

\begin{figure}
\epsscale{1.2}
\plotone{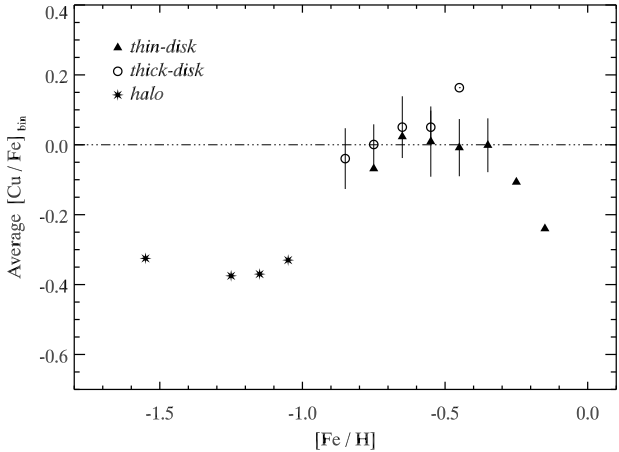}
\caption{The trend of average [Cu/Fe] for 0.1 dex bin of [Fe/H], only non-LTE results are presented. Stars from different populations are averaged separately, and the peculiar stars are not included. The error bars represent the abundance dispersions of the corresponding bins.\\ \label{fig7}}
\end{figure}

Figure \ref{fig3} displays the results of [Cu/Fe] in our program stars as a function of [Fe/H] for both LTE and non-LTE calculations. The [Cu/Fe] trend is similar to the earlier work for the LTE results, while it is not the case for non-LTE ones. In order to show the features more clearly, we present the average [Cu/Fe] values for each $0.1$ dex bin of [Fe/H] ($\Delta$[Fe/H] $=0.1$ bin) in Figure \ref{fig7} for non-LTE results. Stars from different populations are averaged separately, and the peculiar stars are not included. The error bars represent the abundance dispersions of the corresponding bins (no bar is plotted if there is only one or two stars in the bin). A flat distribution of [Cu/Fe] can be seen for our non-LTE results in the range of $-1.5 <$ [Fe/H] $< -1.0$, which is different from that revealed by the previous work (a linear increase from [Fe/H] $= -1.5$ to $-1.0$). However, it should be noted that there are only four stars in that region, and more data are needed to confirm this trend. Even though there may exist bias, the [Cu/Fe] trend derived from non-LTE is still much flatter than that from LTE at $-1.5 <$ [Fe/H] $< -1.0$. The [Cu/Fe] trend for disk stars is not discussed often. \citet{Pro00} studied the copper abundances for $10$ thick-disk stars, and a supersolar [Cu/Fe] can be seen in their results at [Fe/H] $\sim -0.4$. \citet{Red06} suggested that the [Cu/Fe] seemed slightly greater for thick-disk stars than that for thin-disk stars. Our results indicate that the [Cu/Fe] gradually increases with the increasing [Fe/H] for the thick-disk stars, while most of the thin-disk stars have a solar [Cu/Fe] value in the overlapping region. Thus, the thick-disk population appears a slight overabundance of [Cu/Fe] at $-0.7 <$ [Fe/H] $< -0.4$.

Several groups \citep[e.g.,][]{Mat93,Tim95,Mis02,Kob06,Rom07,Rom10} have modelled the Galactic chemical evolution of copper. The most difficult part in the modeling is to approach satisfactory copper abundances at both metal-poor end and solar-metallicity: normal type II supernovae yields from \citet{Woo95}  case B give a good approaching to the observed [Cu/Fe] trend for [Fe/H] $> -2$, but lead to overabundant for copper at lower metallicity, while the yields from \citet{Kob06} SNe II mixed with hypernovae give a better fitting in the metal-poor end but fail to reproduce the observational trend at the solar-metallicity \citep{Rom10}. The ways to solve this inconsistency are so limited because: (1) the contribution from $s-$process cannot be changed freely \citep{Mat93} and (2) there are still large uncertainties in modeling the yields from low- and intermediate-mass AGB stars or supernovae explosion events \citep{Rom07,Rom10}. Since the departures from LTE for Cu I show a clear dependence on metallicity, the copper abundances of the very/extremely metal-poor stars are expected to increase towards lower metallicity (\citeauthor{Roe14} \citeyear{Roe14}; Shi et al. 2014, in preparation). Consequently, before modifying GCE models, a firm and reliable observational trend of [Cu/Fe] should be established first.

\subsection{the Bending} \label{sec6-2}

The non-LTE results for disk stars suggest that there may be a bending-like feature at [Fe/H] $> -1.0$ (Figure \ref{fig3}), where the [Cu/Fe] trend goes up at first, but slightly decreases as the [Fe/H] becomes higher. The feature can be seen in Figure \ref{fig7} as well. The term `bending' was introduced by \citet{Bis04,Bis05} and \citet{Rom07}. Those authors superimposed the [Cu/Fe] data from different work, and a bending for disk stars appeared in the overlapping region, however, the possibility that the bending may be caused by the systematic offset between different studies could not be ruled out. The similar feature can be found for our non-LTE results, and it should be noted that we do not have enough data at [Fe/H] $> -0.3$, the decline in this region is mainly produced by three stars, and as a result, more data are needed to draw a firm conclusion.

 One of the explanations for the bending might be SNe Ia. However, the typical timescale for the chemical enrichment from SNe Ia in the Milky Way is model-dependent \citep[e.g.,][]{Mat01} and thus not strictly constrained.

\section{CONCLUSIONS}

We have investigated the copper abundances for $64$ late-type stars with effective temperatures from $5400$ K to $6700$ K and [Fe/H] between $-1.88$ and $-0.17$. The non-LTE statistical equilibrium calculations are performed to derive the copper abundances. Based on our results, we come to the following conclusions.

\begin{enumerate}
\item The non-LTE effects are significantly large for copper, and they differ from line to line. Two \ion{Cu}{1} lines, $5105$\,\AA\ and $5782$\,\AA\ are more sensitive to non-LTE effects. The non-LTE corrections for these lines can reach $\sim 0.2$ dex at the metallicity [Fe/H] $\sim -1.5$. The weaker line $5218$\,\AA\ shows relatively small but still not negligible non-LTE effects, with a maximum departure of $0.13$ dex in our sample.

\item The copper abundances are underestimated in LTE calculations. Taking non-LTE effects into account, the copper abundances increase for all of our program stars.

\item The non-LTE effects clearly show dependence on metallicity, and they increase with decreasing [Fe/H].

\item Our non-LTE results show that there might be a [Cu/Fe] plateau in the metallicity range $-1.5 <$ [Fe/H] $< -1.0$, however, it need more data to confirm this result. There is a hint that the thick- and thin-disk stars have different behaviors in [Cu/Fe], and a bending for disk stars may exist.
\end{enumerate}

Further non-LTE studies to copper abundance in various environments with wider metallicity range need to be carried out. They are essential to the understanding of the copper origination and evolution.

\acknowledgments
This research was supported by the National Natural Science Foundation of China under grant Nos. 11321064, 11233004, 11390371, 11473033, U1331122, and by National Key Basic Research Program of China 2014CB845700.

\clearpage
\begin{table*}
\begin{center}
\caption[1]{Copper abundances of our program stars \label{tab2}}
\begin{tabular}{lrrrrrrrrrr}
\hline
\hline
\noalign{\smallskip}
Star & $\Teff$ & $\log g$ & [Fe/H] & $\xi$ & $5105$\AA & $5218$\AA & $5782$\AA & [Cu/Fe] & POP & LinFor\\
\noalign{\smallskip}
\hline
\noalign{\smallskip}
 HD\,$            17948$ & $6325$ & $4.13$ & $-0.35$ & $1.90$ & $-0.25$ & $-0.19$ & $-0.16$ & $-0.20\pm0.05$ & D & L \\
                         &        &        &         &        & $-0.18$ & $-0.13$ & $-0.09$ & $-0.13\pm0.05$ &   & N \\
 HD\,$            22309$ & $5900$ & $4.29$ & $-0.31$ & $1.30$ & $ 0.03$ & $-0.04$ & $-0.03$ & $-0.01\pm0.04$ & D & L \\
                         &        &        &         &        & $ 0.07$ & $-0.02$ & $ 0.01$ & $ 0.02\pm0.05$ &   & N \\
 HD\,$            22879$ & $5775$ & $4.26$ & $-0.83$ & $1.10$ & $-0.25$ & $-0.13$ & $-0.15$ & $-0.18\pm0.06$ & T & L \\
                         &        &        &         &        & $-0.17$ & $-0.10$ & $-0.08$ & $-0.12\pm0.05$ &   & N \\
 HD\,$            30649$ & $5765$ & $4.26$ & $-0.58$ & $1.10$ & $ 0.03$ & $ 0.03$ & $ 0.02$ & $ 0.03\pm0.01$ & T & L \\
                         &        &        &         &        & $ 0.09$ & $ 0.05$ & $ 0.07$ & $ 0.07\pm0.02$ &   & N \\
 HD\,$           243357$ & $5675$ & $4.38$ & $-0.59$ & $1.10$ & $-0.04$ & $ 0.05$ & $-0.02$ & $ 0.00\pm0.05$ & T & L \\
                         &        &        &         &        & $ 0.01$ & $ 0.07$ & $ 0.03$ & $ 0.04\pm0.03$ &   & N \\
 HD\,$            36283$ & $5475$ & $4.28$ & $-0.41$ & $0.80$ & $ 0.01$ & $ 0.13$ & $ 0.05$ & $ 0.06\pm0.06$ & T & L \\
                         &        &        &         &        & $ 0.04$ & $ 0.13$ & $ 0.08$ & $ 0.08\pm0.05$ &   & N \\
  G\,$            99-21$ & $5525$ & $4.30$ & $-0.63$ & $1.00$ & $-0.13$ & $ 0.01$ & $-0.10$ & $-0.07\pm0.07$ & T & L \\
                         &        &        &         &        & $-0.09$ & $ 0.03$ & $-0.06$ & $-0.04\pm0.06$ &   & N \\
 HD\,$           250792$ & $5600$ & $4.32$ & $-1.02$ & $1.10$ & $-0.38$ & $-0.40$ & $-0.38$ & $-0.39\pm0.01$ & H & L \\
                         &        &        &         &        & $-0.30$ & $-0.38$ & $-0.31$ & $-0.33\pm0.04$ &   & N \\
 HD\,$            46341$ & $5880$ & $4.36$ & $-0.58$ & $1.80$ & $-0.05$ & $-0.07$ & $-0.08$ & $-0.07\pm0.02$ & T & L \\
                         &        &        &         &        & $ 0.01$ & $-0.04$ & $-0.02$ & $-0.02\pm0.03$ &   & N \\
 HD\,$            56513$ & $5630$ & $4.53$ & $-0.45$ & $1.20$ & $-0.06$ & $-0.01$ & $ 0.01$ & $-0.02\pm0.04$ & D & L \\
                         &        &        &         &        & $-0.03$ & $ 0.00$ & $ 0.04$ & $ 0.00\pm0.04$ &   & N \\
 HD\,$            58551$ & $6190$ & $4.23$ & $-0.53$ & $1.80$ & $-0.13$ & $-0.07$ & $-0.02$ & $-0.07\pm0.06$ & D & L \\
                         &        &        &         &        & $-0.05$ & $-0.01$ & $ 0.06$ & $ 0.00\pm0.06$ &   & N \\
 HD\,$            59374$ & $5840$ & $4.37$ & $-0.83$ & $1.40$ & $-0.16$ & $-0.17$ & $-0.08$ & $-0.14\pm0.05$ & T & L \\
                         &        &        &         &        & $-0.08$ & $-0.13$ & $ 0.00$ & $-0.07\pm0.07$ &   & N \\
 HD\,$            59984$ & $5925$ & $3.94$ & $-0.74$ & $1.20$ & $-0.12$ & $-0.06$ & $-0.08$ & $-0.09\pm0.03$ & T & L \\
                         &        &        &         &        & $-0.02$ & $ 0.00$ & $ 0.01$ & $ 0.00\pm0.02$ &   & N \\
 HD\,$            60319$ & $5875$ & $4.17$ & $-0.82$ & $1.40$ & $-0.15$ & $-0.10$ & $-0.15$ & $-0.13\pm0.03$ & T & L \\
                         &        &        &         &        & $-0.06$ & $-0.05$ & $-0.06$ & $-0.06\pm0.01$ &   & N \\
  G\,$           235-45$ & $5500$ & $4.25$ & $-0.59$ & $1.10$ & $ 0.01$ & $-0.01$ & $ 0.05$ & $ 0.02\pm0.03$ & T & L \\
                         &        &        &         &        & $ 0.05$ & $ 0.00$ & $ 0.09$ & $ 0.05\pm0.05$ &   & N \\
 HD\,$            88446$ & $5915$ & $4.03$ & $-0.44$ & $1.60$ & $-0.01$ & $-0.08$ & $-0.14$ & $-0.08\pm0.07$ & D & L \\
                         &        &        &         &        & $-0.04$ & $-0.04$ & $-0.07$ & $-0.05\pm0.02$ &   & N \\
 HD\,$            88725$ & $5665$ & $4.35$ & $-0.70$ & $1.20$ & $-0.01$ & $ 0.03$ & $ 0.03$ & $ 0.02\pm0.02$ & T & L \\
                         &        &        &         &        & $ 0.05$ & $ 0.06$ & $ 0.09$ & $ 0.07\pm0.02$ &   & N \\
 HD\,$            91784$ & $5890$ & $4.47$ & $-0.33$ & $1.30$ & $-0.01$ & $ 0.06$ & $ 0.06$ & $ 0.04\pm0.04$ & D & L \\
                         &        &        &         &        & $ 0.03$ & $ 0.08$ & $ 0.10$ & $ 0.07\pm0.04$ &   & N \\
 HD\,$            94028$ & $5925$ & $4.19$ & $-1.51$ & $1.50$ & $-0.57$ & $-0.41$ & \nodata & $-0.49\pm0.11$ & H & L \\
                         &        &        &         &        & $-0.37$ & $-0.28$ & \nodata & $-0.32\pm0.06$ &   & N \\
 HD\,$            96094$ & $5900$ & $4.01$ & $-0.46$ & $1.70$ & $-0.05$ & $-0.06$ & $ 0.01$ & $-0.03\pm0.04$ & D & L \\
                         &        &        &         &        & $ 0.02$ & $-0.02$ & $ 0.07$ & $ 0.02\pm0.05$ &   & N \\
 HD\,$           97855A$ & $6240$ & $4.13$ & $-0.44$ & $1.80$ & $-0.22$ & $-0.15$ & $-0.11$ & $-0.16\pm0.06$ & D & L \\
                         &        &        &         &        & $-0.14$ & $-0.09$ & $-0.03$ & $-0.09\pm0.06$ &   & N \\
 HD\,$           101177$ & $5890$ & $4.30$ & $-0.47$ & $1.80$ & $ 0.09$ & $ 0.04$ & $ 0.14$ & $ 0.09\pm0.05$ & D & L \\
                         &        &        &         &        & $ 0.16$ & $ 0.07$ & $ 0.20$ & $ 0.14\pm0.07$ &   & N \\
 HD\,$           104056$ & $5875$ & $4.31$ & $-0.41$ & $1.30$ & $-0.02$ & $ 0.00$ & $ 0.02$ & $ 0.00\pm0.02$ & D & L \\
                         &        &        &         &        & $ 0.03$ & $ 0.02$ & $ 0.06$ & $ 0.04\pm0.02$ &   & N \\
 HD\,$           107582$ & $5565$ & $4.34$ & $-0.61$ & $1.00$ & $-0.10$ & $-0.09$ & $-0.15$ & $-0.11\pm0.03$ & T & L \\
                         &        &        &         &        & $-0.06$ & $-0.07$ & $-0.11$ & $-0.08\pm0.03$ &   & N \\
 HD\,$           108076$ & $5725$ & $4.44$ & $-0.73$ & $1.20$ & $-0.20$ & $-0.17$ & $-0.21$ & $-0.19\pm0.02$ & D & L \\
                         &        &        &         &        & $-0.14$ & $-0.15$ & $-0.16$ & $-0.15\pm0.01$ &   & N \\
 HD\,$           114606$ & $5610$ & $4.28$ & $-0.57$ & $1.20$ & $ 0.04$ & $ 0.06$ & $ 0.02$ & $ 0.04\pm0.02$ & T & L \\
                         &        &        &         &        & $ 0.09$ & $ 0.08$ & $ 0.06$ & $ 0.08\pm0.02$ &   & N \\
 HD\,$           118659$ & $5510$ & $4.36$ & $-0.60$ & $1.00$ & $-0.03$ & $ 0.01$ & $ 0.02$ & $ 0.00\pm0.03$ & T & L \\
                         &        &        &         &        & $ 0.02$ & $ 0.02$ & $ 0.06$ & $ 0.03\pm0.02$ &   & N \\
 HD\,$           119288$ & $6420$ & $4.13$ & $-0.17$ & $1.90$ & $-0.34$ & \nodata & $-0.26$ & $-0.30\pm0.06$ & D & L \\
                         &        &        &         &        & $-0.28$ & \nodata & $-0.20$ & $-0.24\pm0.06$ &   & N \\
\noalign{\smallskip}
\hline
\noalign{\smallskip}
\end{tabular}
\end{center}
\textbf{Notes.} Both LTE and non-LTE copper abundances (for each star, first and second row, respectively) of our program stars are listed in column $9$. The abundances given here are the relative values to [Fe/H] derived from \ion{Fe}{2} lines (Section \ref{sec3-2}). Column $6$, $7$, and $8$ are the abundances derived from corresponding \ion{Cu}{1} lines. The stellar parameters and population assignments are also shown in the table. The characters `D', `T', `H' and `?' in the 'POP' column represent thin-disk, thick-disk, halo, and peculiar stars, respectively. The rightmost column indicates the line formation scenario for each star, where 'L' represents LTE line formation and 'N' represents non-LTE line formation.\\
\end{table*}

\begin{table*}
\begin{center}
\tablename\ \thetable{} -- continued \\ [0.1cm]
\begin{tabular}{lrrrrrrrrrr}
\hline
\hline
\noalign{\smallskip}
Star & $\Teff$ & $\log g$ & [Fe/H] & $\xi$ & $5105$\AA & $5218$\AA & $5782$\AA & [Cu/Fe] & Pop & LinFor \\
\noalign{\smallskip}
\hline
\noalign{\smallskip}
 HD\,$           123710$ & $5790$ & $4.41$ & $-0.54$ & $1.40$ & $ 0.07$ & $ 0.08$ & $ 0.06$ & $ 0.07\pm0.01$ & D & L \\
                         &        &        &         &        & $ 0.13$ & $ 0.11$ & $ 0.11$ & $ 0.12\pm0.01$ &   & N \\
 HD\,$           126512$ & $5825$ & $4.02$ & $-0.64$ & $1.60$ & $-0.03$ & $ 0.02$ & $ 0.06$ & $ 0.02\pm0.05$ & T & L \\
                         &        &        &         &        & $ 0.05$ & $ 0.07$ & $ 0.14$ & $ 0.09\pm0.05$ &   & N \\
 HD\,$           134169$ & $5930$ & $3.98$ & $-0.86$ & $1.80$ & $ 0.01$ & $-0.04$ & \nodata & $-0.02\pm0.04$ & T & L \\
                         &        &        &         &        & $ 0.13$ & $ 0.04$ & \nodata & $ 0.09\pm0.06$ &   & N \\
 HD\,$           142267$ & $5807$ & $4.42$ & $-0.46$ & $1.00$ & $ 0.01$ & $-0.06$ & $-0.02$ & $-0.02\pm0.04$ & D & L \\
                         &        &        &         &        & $ 0.06$ & $-0.04$ & $ 0.02$ & $ 0.01\pm0.05$ &   & N \\
 HD\,$           144061$ & $5815$ & $4.44$ & $-0.31$ & $1.20$ & $ 0.00$ & $-0.06$ & $ 0.04$ & $-0.01\pm0.05$ & D & L \\
                         &        &        &         &        & $ 0.04$ & $-0.05$ & $ 0.07$ & $ 0.02\pm0.06$ &   & N \\
 HD\,$           148816$ & $5880$ & $4.07$ & $-0.78$ & $1.20$ & $-0.04$ & $-0.07$ & $-0.02$ & $-0.04\pm0.03$ & ? & L \\
                         &        &        &         &        & $ 0.06$ & $-0.02$ & $ 0.07$ & $ 0.04\pm0.05$ &   & N \\
 HD\,$           149996$ & $5665$ & $4.09$ & $-0.52$ & $1.20$ & $ 0.00$ & $ 0.08$ & $ 0.00$ & $ 0.03\pm0.05$ & T & L \\
                         &        &        &         &        & $ 0.06$ & $ 0.11$ & $ 0.05$ & $ 0.07\pm0.03$ &   & N \\
 BD\,$ +68^{\circ}\,901$ & $5715$ & $4.51$ & $-0.25$ & $1.40$ & $-0.12$ & $-0.14$ & $-0.04$ & $-0.10\pm0.05$ & D & L \\
                         &        &        &         &        & $-0.09$ & $-0.13$ & $-0.01$ & $-0.08\pm0.06$ &   & N \\
 HD\,$           157089$ & $5800$ & $4.06$ & $-0.59$ & $1.20$ & $-0.08$ & $-0.09$ & $-0.05$ & $-0.07\pm0.02$ & T & L \\
                         &        &        &         &        & $-0.01$ & $-0.06$ & $ 0.01$ & $-0.02\pm0.04$ &   & N \\
 HD\,$           157466$ & $5990$ & $4.38$ & $-0.44$ & $1.10$ & $-0.20$ & $-0.22$ & $-0.15$ & $-0.19\pm0.04$ & D & L \\
                         &        &        &         &        & $-0.14$ & $-0.20$ & $-0.10$ & $-0.15\pm0.05$ &   & N \\
 HD\,$           158226$ & $5805$ & $4.12$ & $-0.56$ & $1.10$ & $ 0.04$ & $-0.03$ & $ 0.04$ & $ 0.02\pm0.04$ & T & L \\
                         &        &        &         &        & $ 0.11$ & $ 0.00$ & $ 0.10$ & $ 0.07\pm0.06$ &   & N \\
  G\,$           170-56$ & $6030$ & $4.31$ & $-0.79$ & $1.30$ & $-0.42$ & $-0.29$ & \nodata & $-0.35\pm0.09$ & ? & L \\
                         &        &        &         &        & $-0.33$ & $-0.24$ & \nodata & $-0.28\pm0.06$ &   & N \\
 HD\,$           160933$ & $5765$ & $3.85$ & $-0.27$ & $1.20$ & $-0.16$ & $-0.19$ & $-0.16$ & $-0.17\pm0.02$ & D & L \\
                         &        &        &         &        & $-0.12$ & $-0.17$ & $-0.12$ & $-0.14\pm0.03$ &   & N \\
 HD\,$           160693$ & $5850$ & $4.31$ & $-0.60$ & $1.20$ & $ 0.05$ & $ 0.01$ & $ 0.08$ & $ 0.05\pm0.04$ & ? & L \\
                         &        &        &         &        & $ 0.12$ & $ 0.04$ & $ 0.14$ & $ 0.10\pm0.05$ &   & N \\
 HD\,$           170357$ & $5665$ & $4.07$ & $-0.50$ & $1.20$ & $-0.03$ & $-0.04$ & $ 0.03$ & $-0.01\pm0.04$ & T & L \\
                         &        &        &         &        & $ 0.02$ & $-0.02$ & $ 0.08$ & $ 0.03\pm0.05$ &   & N \\
 HD\,$           171620$ & $6115$ & $4.20$ & $-0.50$ & $1.40$ & $-0.02$ & $-0.06$ & $ 0.01$ & $-0.02\pm0.04$ & D & L \\
                         &        &        &         &        & $ 0.06$ & $-0.01$ & $ 0.08$ & $ 0.04\pm0.05$ &   & N \\
  G\,$            142-2$ & $5675$ & $4.48$ & $-0.75$ & $1.10$ & $-0.03$ & $-0.02$ & $-0.04$ & $-0.03\pm0.01$ & T & L \\
                         &        &        &         &        & $ 0.03$ & $ 0.00$ & $ 0.01$ & $ 0.01\pm0.02$ &   & N \\
 HD\,$           182807$ & $6100$ & $4.21$ & $-0.33$ & $1.40$ & $-0.04$ & $-0.06$ & $ 0.00$ & $-0.03\pm0.03$ & D & L \\
                         &        &        &         &        & $ 0.02$ & $-0.03$ & $ 0.06$ & $ 0.02\pm0.05$ &   & N \\
 HD\,$           184448$ & $5765$ & $4.16$ & $-0.43$ & $1.20$ & $ 0.21$ & $ 0.17$ & $ 0.22$ & $ 0.20\pm0.03$ & T & L \\
                         &        &        &         &        & $ 0.27$ & $ 0.19$ & $ 0.27$ & $ 0.24\pm0.05$ &   & N \\
 HD\,$           186379$ & $5865$ & $3.93$ & $-0.41$ & $1.20$ & $-0.06$ & $-0.06$ & $-0.06$ & $-0.06\pm0.00$ & D & L \\
                         &        &        &         &        & $ 0.00$ & $-0.03$ & $ 0.00$ & $-0.01\pm0.02$ &   & N \\
 HD\,$           198300$ & $5890$ & $4.31$ & $-0.60$ & $1.20$ & $ 0.09$ & $ 0.02$ & $ 0.06$ & $ 0.06\pm0.04$ & T & L \\
                         &        &        &         &        & $ 0.16$ & $ 0.06$ & $ 0.12$ & $ 0.11\pm0.05$ &   & N \\
 HD\,$           200580$ & $5940$ & $3.96$ & $-0.82$ & $1.40$ & $ 0.16$ & $ 0.18$ & $ 0.28$ & $ 0.21\pm0.06$ & ? & L \\
                         &        &        &         &        & $ 0.28$ & $ 0.26$ & $ 0.39$ & $ 0.31\pm0.07$ &   & N \\
  G\,$           188-22$ & $6040$ & $4.37$ & $-1.25$ & $1.50$ & $-0.54$ & $-0.46$ & \nodata & $-0.50\pm0.06$ & H & L \\
                         &        &        &         &        & $-0.39$ & $-0.36$ & \nodata & $-0.38\pm0.02$ &   & N \\
 HD\,$           201889$ & $5710$ & $4.05$ & $-0.78$ & $1.10$ & $-0.11$ & $-0.17$ & $-0.13$ & $-0.14\pm0.03$ & T & L \\
                         &        &        &         &        & $-0.03$ & $-0.13$ & $-0.06$ & $-0.07\pm0.05$ &   & N \\
 HD\,$           204155$ & $5815$ & $4.09$ & $-0.66$ & $1.20$ & $-0.02$ & $ 0.00$ & $ 0.02$ & $ 0.00\pm0.02$ & T & L \\
                         &        &        &         &        & $ 0.06$ & $ 0.04$ & $ 0.09$ & $ 0.06\pm0.03$ &   & N \\
 HD\,$           208906$ & $6025$ & $4.37$ & $-0.76$ & $1.40$ & $-0.07$ & $-0.09$ & $-0.02$ & $-0.06\pm0.04$ & D & L \\
                         &        &        &         &        & $ 0.02$ & $-0.04$ & $ 0.06$ & $ 0.01\pm0.05$ &   & N \\
  G\,$            242-4$ & $5815$ & $4.31$ & $-1.10$ & $1.20$ & $-0.52$ & $-0.44$ & $-0.40$ & $-0.45\pm0.06$ & H & L \\
                         &        &        &         &        & $-0.42$ & $-0.39$ & $-0.30$ & $-0.37\pm0.06$ &   & N \\
 HD\,$           215257$ & $6030$ & $4.28$ & $-0.58$ & $1.40$ & $-0.18$ & $-0.22$ & $-0.14$ & $-0.18\pm0.04$ & D & L \\
                         &        &        &         &        & $-0.11$ & $-0.18$ & $-0.08$ & $-0.12\pm0.05$ &   & N \\
 HD\,$           218209$ & $5665$ & $4.40$ & $-0.60$ & $1.10$ & $ 0.14$ & $ 0.09$ & $ 0.17$ & $ 0.13\pm0.04$ & T & L \\
                         &        &        &         &        & $ 0.20$ & $ 0.11$ & $ 0.22$ & $ 0.18\pm0.06$ &   & N \\
 HD\,$           221876$ & $5865$ & $4.29$ & $-0.60$ & $1.20$ & $-0.07$ & $-0.07$ & $ 0.05$ & $-0.03\pm0.07$ & D & L \\
                         &        &        &         &        & $ 0.00$ & $-0.04$ & $ 0.11$ & $ 0.02\pm0.08$ &   & N \\
 HD\,$           224930$ & $5480$ & $4.45$ & $-0.66$ & $0.90$ & $-0.16$ & $-0.16$ & $-0.15$ & $-0.16\pm0.01$ & ? & L \\
                         &        &        &         &        & $-0.12$ & $-0.15$ & $-0.12$ & $-0.13\pm0.02$ &   & N \\
  G\,$             69-8$ & $5640$ & $4.43$ & $-0.55$ & $1.10$ & $ 0.13$ & $ 0.07$ & $ 0.11$ & $ 0.10\pm0.03$ & T & L \\
                         &        &        &         &        & $ 0.18$ & $ 0.09$ & $ 0.15$ & $ 0.14\pm0.05$ &   & N \\
\noalign{\smallskip}
\hline
\noalign{\smallskip}
\end{tabular}
\end{center}
\end{table*}

\end{document}